\documentclass{article}

\usepackage[final]{neurips_2025}

\usepackage[utf8]{inputenc} 
\usepackage[T1]{fontenc}    
\usepackage{hyperref}       
\usepackage{url}            
\usepackage{booktabs}       
\usepackage{amsfonts}       
\usepackage{nicefrac}       
\usepackage{microtype}      
\usepackage{xcolor}         
\usepackage{graphicx}

\usepackage{array}    

\title{Finance as Extended Biology: Reciprocity as the Cognitive Substrate of Financial Behavior}

\author{%
  Egil Diau \\
  Department of Computer Science\\
  Nation Taiwan University\\
  Taiwan, Taipei \\
  \texttt{egil158@gmail.com} \\
}

\begin{document}

\maketitle

\begin{abstract}
A central challenge in economics and artificial intelligence is explaining how financial behaviors—such as credit, insurance, and trade—emerge without formal institutions. We argue that these functions are not products of institutional design, but structured extensions of a single behavioral substrate: reciprocity. Far from being a derived strategy, reciprocity served as the foundational logic of early human societies—governing the circulation of goods, regulation of obligation, and maintenance of long-term cooperation well before markets, money, or formal rules. Trade, commonly regarded as the origin of financial systems, is reframed here as the canonical form of reciprocity: simultaneous, symmetric, and partner-contingent. Building on this logic, we reconstruct four core financial functions—credit, insurance, token exchange, and investment—as expressions of the same underlying principle under varying conditions. By grounding financial behavior in minimal, simulateable dynamics of reciprocal interaction, this framework shifts the focus from institutional engineering to behavioral computation—offering a new foundation for modeling decentralized financial behavior in both human and artificial agents.
\end{abstract}

\begin{figure}[htbp]
    \centering
    \includegraphics[width=0.8\textwidth]{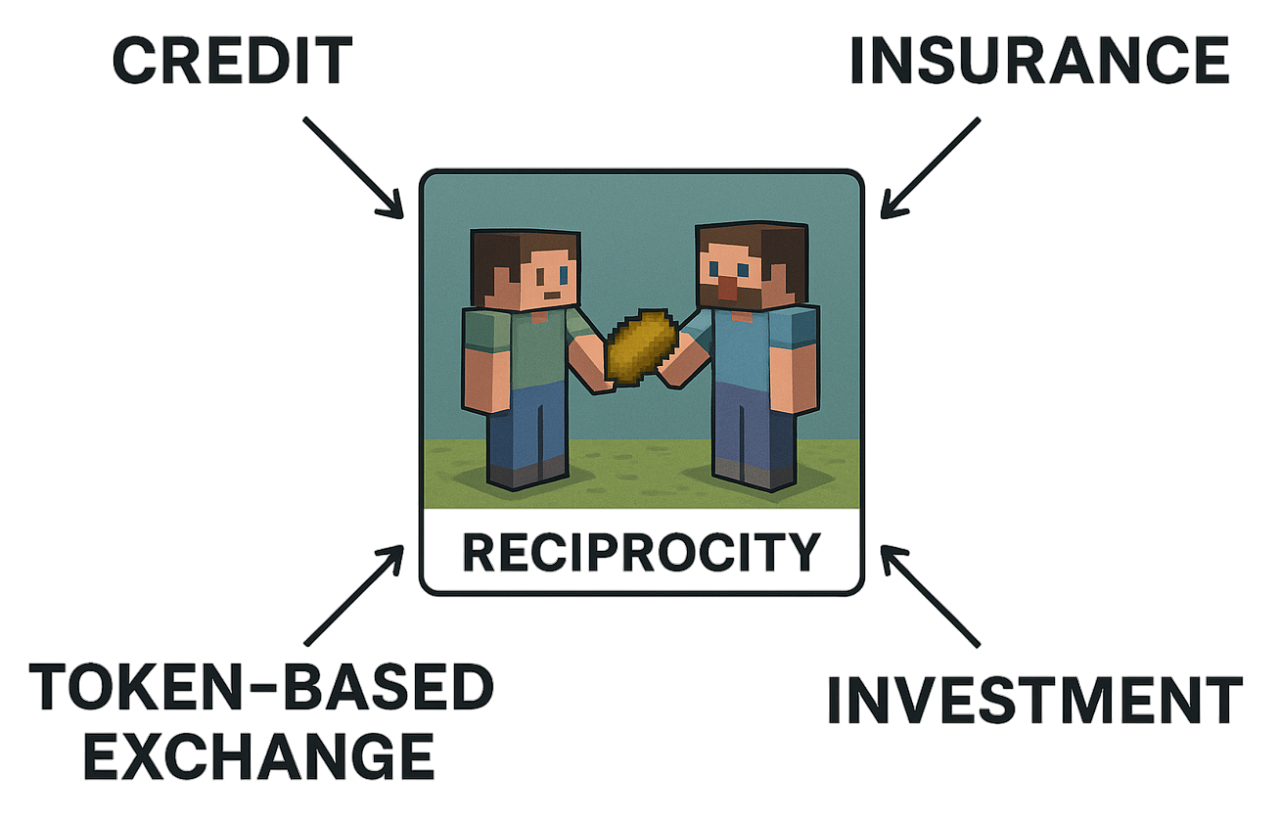}
    \caption{Reciprocity underlies trade and its structural extensions—credit, insurance, token exchange, and investment—each shaped by delay, risk, indirection, and future reward.}
    \label{fig:main}
\end{figure}

\section{Introduction}
Many core financial behaviors—credit, trade, and risk pooling—are typically explained through institutions: contracts, incentives, and enforcement systems. Yet such behaviors appear in small-scale human societies and even non-human primates, long before the advent of formal markets or symbolic money. This raises a fundamental question: can financial systems emerge from social interaction alone?

Most economic models approach finance from the top down—assuming rational agents, institutional scaffolding, and predefined payoff structures. Financial behaviors are often treated as primitives, not reconstructed from minimal cognitive or behavioral mechanisms. But if cooperation is grounded in social cognition, then institutions may be outcomes—not prerequisites—of financial organization.

This limitation extends to simulation. Despite growing interest in emergent economies, multi-agent systems still struggle to reproduce finance-like behavior from the bottom up. Such dynamics are often hardcoded or engineered directly, without identifying the minimal conditions under which they arise. We lack simulateable frameworks showing how minimally capable agents, interacting locally and without global structure, can give rise to financial dynamics through decentralized interaction.

We address this gap by starting from reciprocity—a biologically grounded mechanism for contingent cooperation observed across species \cite{de1997chimpanzee}. Far from peripheral, reciprocity served as the central organizing principle of early human societies, governing exchange, obligation, and long-term relational stability \cite{sahlins2013stone, malinowski2013argonauts, mauss2024gift}. Its robustness and minimal cognitive demands make it a plausible substrate for financial behavior. Crucially, it requires no formal rules or enforcement, allowing reconstruction from first behavioral principles.

Yet while reciprocity is well studied in anthropology and behavioral science, it is rarely formalized in economic or computational models. Trade—widely regarded as the foundation of finance—is typically framed as a distinct economic behavior. But its structure—simultaneous, symmetric, and partner-contingent—suggests it is simply the most basic form of reciprocity. Recognizing this reframes trade not as a separate mechanism, but as the baseline case of contingent cooperation. Other financial behaviors—credit, insurance, token exchange, and investment—can then be understood as structured extensions of the same process.

Building on this insight, we extend a recently proposed framework for simulateable reciprocity \cite{diau2025cognitivefoundationseconomicexchange} to reconstruct four foundational financial functions as structural transformations of a single behavioral substrate: reciprocity. Specifically, we frame:

\begin{enumerate}
    \item \textbf{Credit}: reciprocity under time delay;
    \item \textbf{Insurance}: reciprocity under uncertainty and asymmetric risk;
    \item \textbf{Token-based exchange}: reciprocity mediated through indirect links;
    \item \textbf{Investment}: reciprocity oriented toward expected future reward.
\end{enumerate}

These functions are not independent mechanisms, but emergent structures shaped by distinct constraints. Each is grounded in minimal agent capacities: partner recognition, memory of prior interaction, and cost–return sensitivity. Rather than treating financial institutions as designed rules, we view them as dynamic macrostates—stabilized through recursive interaction and observable through simulation.

\paragraph{Our contribution.}
This work offers a theoretical realignment: a behavioral reconstruction of financial structure. We:

\begin{itemize}
    \item Demonstrate that trade, often attributed to institutional design, is a canonical expression of reciprocity;
    \item Reframe four core financial functions—credit, insurance, token exchange, and investment—as structural extensions of the same underlying mechanism;
    \item Define a minimal, simulateable agent substrate capable of supporting these behaviors without institutional scaffolding;
    \item Provide a framework for modeling decentralized financial emergence in both economic and artificial agent systems.
\end{itemize}

\paragraph{Ethical Statement}
This work does not rely on evolutionary explanations or biological determinism.
While we draw on behavioral evidence from primates and early human societies, our goal is not to claim innate or adaptive origins of social structures.
Rather, these observations serve as empirical constraints to inform the design of minimal cognitive and behavioral mechanisms sufficient for the emergence of reciprocal norms and institutions.
Our account focuses on behavioral plausibility and simulateable processes, without assuming evolutionary teleology, cultural essentialism, or autonomous social evolution.

\section{Related Work}
\label{gen_inst}

\subsection{Agent-Based Modeling of Social Behaviors}

Most agent-based economic models simulate financial behavior by assuming institutional structures—markets, contracts, or explicit payoff matrices. Agents are typically assigned predefined economic roles (e.g., buyer, lender, insurer) and interact under engineered incentive rules. While such models can produce surface-level economic regularities, they rarely account for how structured financial behaviors emerge from the interaction dynamics themselves.

Recent work in multi-agent reinforcement learning and language model-based systems has demonstrated the capacity for negotiation, cooperation, and resource sharing \citep{park2023generative, li2023camel}. However, these simulations often operate at the level of abstract policies or token exchanges, with little attention to the behavioral logic that underlies real-world financial behavior—such as trust formation, delayed obligation, or indirect exchange.

As a result, current simulations tend to treat financial systems as given environments rather than as structures emergent from interactions. The question of how core financial functions such as credit, insurance, and investment can spontaneously arise from minimal, repeated interactions remains underexplored in existing literature.

\subsection{Behavioral Foundations of Reciprocity: Insights from Primatology and Anthropology}

Primate research offers foundational insight into the behavioral roots of reciprocity. 
Species such as chimpanzees and bonobos exhibit contingent prosocial behaviors across contexts including food sharing, grooming, and coalitionary support. 
Classic work by de Waal \cite{de1997chimpanzee} documented long-term social bonds maintained through reciprocal acts, while Brosnan et al. \cite{brosnan2003monkeys} showed that non-human primates respond negatively to unequal outcomes. 
These observations suggest that reciprocity is not merely a cultural construct, but a broadly observed behavioral pattern across species.

Ethnographic studies of early human societies emphasize reciprocity as the foundational structure of economic life. 
Rather than operating through explicit transactions or market pricing, these communities organized production, distribution, and obligation through systems of delayed return, social debt, and mutual obligation \cite{mauss2024gift, sahlins2013stone, malinowski2013argonauts}. 
Reciprocity was not only an economic practice, but the principle that sustained cooperation, structured relationships, and enabled long-term coordination within and across groups.

If reciprocity constitutes one of the earliest recognizable forms of economic behavior, then finance is not a categorical departure—but a structured extension.
Trade, often regarded as the foundation of finance, can be reframed as a formalized instance of reciprocal interaction: symmetrical, temporally synchronized, and partner-contingent.
When reciprocity operates under distinct structural conditions—such as temporal delay, risk asymmetry, indirect mediation, or future-oriented return—it gives rise to four core financial functions: credit, insurance, token-based exchange, and investment.
These functions are best understood not as discrete institutional inventions, but as structural transformations of a shared cooperative substrate.

\subsection{Classical Finance Theories and Behavioral Finance}

Classical financial theory models markets through equilibrium, optimization, and institutional enforcement. Foundational frameworks—including the Efficient Market Hypothesis \cite{fama1970efficient}, Arrow–Debreu Equilibrium \cite{arrow2024existence}, and CAPM \cite{sharpe1964capital}—assume pre-existing financial structures such as credit and trade. These models rely on rational agents interacting within fixed payoff structures.

Behavioral finance introduces bounded rationality and psychological insights through frameworks like Prospect Theory \cite{tversky1992advances}, Mental Accounting \cite{thaler2005advances}, and Adaptive Markets \cite{lo2004adaptive}, emphasizing deviations from rational choice. However, these deviations are modeled within existing market structures, without addressing their behavioral origins. Consequently, the cognitive and social foundations underlying financial behaviors remain insufficiently theorized.

\section{Background: From Financial Theory to Behavioral Foundations}
\label{headings}

\subsection{Traditional Financial Theory: Institutions Without Origins}

Classical finance is built upon models of equilibrium, optimization, and incentive alignment. Foundational frameworks—from general equilibrium theory \cite{arrow2024existence} to the Efficient Market Hypothesis \cite{fama1970efficient} and the Capital Asset Pricing Model \cite{sharpe1964capital}—assume that institutions such as credit, trade, and insurance already exist as formal constructs. These models describe how rational agents behave under defined constraints, but do not explain how those constraints emerge.

Crucially, financial institutions are treated not as outcomes of behavior, but as preconditions for behavior—enforced by legal systems, symbolic contracts, or exogenous markets. As a result, these models offer no account of how decentralized agents, operating without such infrastructure, might spontaneously generate the functional equivalents of lending, pooling, or exchange.

\subsection{The Limits of Behavioral Finance: Corrections Without Foundations}

Behavioral finance introduced a crucial refinement: human agents often deviate from rational choice. This body of work incorporates empirical regularities such as loss aversion, mental accounting, and framing effects \cite{tversky1992advances, thaler2005advances}. It expands our understanding of financial behavior under bounded cognition, but remains anchored within pre-existing institutional frameworks.

These models modify utility functions and incorporate psychological biases, but do not reconstruct the environments in which financial roles arise. Institutions are still treated as given. As a result, behavioral finance improves prediction within existing systems, but cannot account for how those systems emerge in the first place.

\subsection{Toward a Behavioral Substrate: From Correction to Construction}

To move beyond the limitations of modeling finance atop pre-assumed institutional structures, we shift from correction to construction. Rather than adapting existing models to fit observed behavior, we reconstruct financial systems from the bottom up—beginning not with rules, but with behaviors. This requires a substrate that is cognitively plausible, empirically grounded, and simulateable within agent-based systems.

We identify \textit{reciprocity} as the minimal substrate for decentralized financial behavior.  
It is not unique to humans: among non-human primates, reciprocal exchange emerges reliably across social contexts, even without formal rules or enforcement \cite{de1997chimpanzee}.  
In early human societies, reciprocity became more than a behavioral tendency—it served as the organizing logic of economic life.  
It governed how communities exchanged goods, distributed obligations, and maintained long-term coordination.  
Long before markets or money, reciprocity structured economic relationships—not as an idealized norm, but as a documented feature of real-world exchange \cite{mauss2024gift, sahlins2013stone, malinowski2013argonauts}.

When reciprocity operates under distinct structural conditions—such as temporal delay, risk asymmetry, indirect mediation, or future-oriented return—it generates core financial behaviors, not as deliberately designed mechanisms, but as emergent structures of social interaction:
\begin{itemize}
    \item \textbf{Credit}: reciprocity extended over time;
    \item \textbf{Insurance}: reciprocity under asymmetric risk and uncertainty;
    \item \textbf{Token-based exchange}: reciprocity mediated through indirect representation;
    \item \textbf{Investment}: reciprocity oriented toward expected long-term gain.
\end{itemize}

\paragraph{The Cognitive Threshold of Financial Behavior.}
Although reciprocity is broadly observed across species, these structured financial extensions typically demand cognitive capacities—such as abstraction, delayed-return tracking, and social inference—that may exceed the stable range of non-human animals.  
It is not reciprocity itself, but the capacity to stabilize and generalize its extensions across contexts, that marks the cognitive boundary of financial behavior.

\begin{table}[htbp]
\centering
\renewcommand{\arraystretch}{1.4}
\begin{tabular}{@{}p{3cm} p{5cm} p{6.5cm}@{}}
\toprule
\textbf{Financial Function} & \textbf{Reciprocity Extension} & \textbf{Behavioral Logic of Finance} \\
\midrule
Credit & Delayed reciprocation & I help you today, and you return the favor later. \\
Insurance & Mutual support under uncertainty & I help you now, trusting you'll help me if I’m ever in trouble. \\
Token Exchange & Indirect reciprocation via tokens & I help a stranger, and someone else helps me later—with a token to keep track. \\
Investment & Future-oriented reciprocation & I give you something now, expecting you’ll return more in the future. \\
\bottomrule
\end{tabular}
\vspace{0.8em}
\caption{Each financial function reflects a different way of extending reciprocity—adjusting how we give and receive across time, risk, and social context.}
\label{tab:financial-functions}
\end{table}

\section{Core Mechanisms: From Reciprocity to Financial Behavior}

\subsection{Reciprocity + Time Delay: From Social Imbalance to Credit Behavior}

\paragraph{Behavioral Mechanism.}  
Credit begins when one agent gives something now and trusts the other to repay later. This creates a temporary imbalance—what we might call a “social debt.” Over repeated interactions, these one-way acts of giving can become stable expectations: I help you today, you help me tomorrow. This form of delayed reciprocity doesn't need formal rules—just memory and recognition. With them, agents can sustain long-term cooperation, even without enforcement.

\paragraph{Empirical Grounding.}  
Behavioral evidence suggests that time-delayed reciprocity predates formal institutions. In chimpanzees, for instance, grooming and coalitionary support often occur without immediate return—indicating a form of asymmetric helping that relies on memory rather than contract \cite{de1997chimpanzee}. Among humans, the expectation to return favors is similarly robust. Failure to reciprocate triggers psychological discomfort (\textit{indebtedness}) and behavioral compensation (\textit{guilt aversion}) \cite{greenberg1983indebtedness, battigalli2007guilt}, even when no formal obligation exists.

These patterns suggest that a felt imbalance—what we might call a proto-debt—serves as a stabilizing force in delayed cooperation. Repeated trust game experiments \cite{engle2004evolution} show that agents dynamically adjust their return behavior based on prior imbalance, effectively simulating repayment. As such patterns stabilize over time, they may scaffold the emergence of more structured expectations—eventually giving rise to the behavioral logic of credit. In this view, credit is not a distinct invention, but an abstraction layered atop a more ancient sense of social debt.

\subsection{Reciprocity + Risk Buffering: From Shared Vulnerability to Insurance Behavior}

\paragraph{Behavioral Mechanism.}
When misfortune strikes individuals unevenly, reciprocity can adapt from direct repayment to mutual support based on need. In such environments, agents help others during periods of bad luck—not because of past help received, but with the expectation that others will do the same if roles reverse in the future. This form of exchange is not about exact balance, but about stabilizing cooperation under uncertainty through shared risk.

\paragraph{Empirical Grounding.}
Risk-buffering reciprocity is widespread in naturalistic contexts, but remains difficult to reproduce in laboratory experiments. It requires long-term interaction, random adversity, and flexible role reversibility—conditions rarely present in experimental designs. Evidence comes from both field and modeling studies. Among Agta hunter-gatherers, individuals preferentially shared food with those in need—especially parents or the resource-poor—rather than with those who were merely cooperative \cite{smith2019friend}. Agent-based simulations of Maasai livestock exchange under the osotua norm demonstrate that asymmetric giving based on need can stabilize decentralized risk pooling and improve group survival \cite{aktipis2011risk}. While such behaviors are well documented, the cognitive mechanisms that enable conditional giving under uncertainty remain underformalized.

\subsection{Reciprocity + Scalable Mediation: From Dyadic Memory to Indirect Exchange}
\paragraph{Behavioral Mechanism.}
Direct reciprocity usually depends on personal memory—agents remember who helped them and return favors over time. But memory doesn’t scale. As groups grow, maintaining reciprocal relationships with everyone becomes impossible. Token-based mediation solves this: by allowing value or obligation to be represented with portable objects, reciprocity no longer depends on knowing individual histories. Tokens act as shared placeholders for past help, enabling indirect exchange among agents without prior interaction.

\paragraph{Empirical Grounding.}
Token-based exchange relies on intermediary representation: the use of transferable objects to encode value or obligation. This marks a key transition in the scalability of reciprocity—allowing cooperative commitments to decouple from direct dyadic memory and persist through abstraction.

Despite its centrality in modern economic systems, this capacity has received surprisingly little attention in developmental and comparative research. It is rarely isolated in experimental paradigms, likely due to the cognitive and methodological challenges of eliciting two-sided valuation and abstract placeholder use.

Nonetheless, early signs appear across species. Children as young as four spontaneously invent placeholder tokens in cooperative play, and non-human primates have demonstrated the ability to use arbitrary tokens for trade in laboratory settings \cite{chen2006basic}. These findings suggest that token-based exchange draws on general cognitive mechanisms rather than institutional rule-following, enabling reciprocity to scale beyond personal memory and immediate context.

\subsection{Reciprocity + Expected Future Reward: From Projected Return to Investment Behavior}
\paragraph{Behavioral Mechanism.}
Investment is a form of reciprocity where one agent gives up resources now, hoping for a greater return in the future. But unlike direct reciprocity, the benefit is uncertain and depends on others’ future actions. This makes investment a gamble on cooperation: agents must project others’ likely behavior over time and accept short-term cost for potential long-term gain. The act of investing reflects not only patience, but a social expectation—that others will reward contribution in the future.

\paragraph{Empirical Grounding.}
Investment behavior requires the ability to accept immediate cost in pursuit of uncertain, delayed gain. This intertemporal capacity emerges early in development, as shown by the marshmallow test, where children delay gratification for a larger reward \cite{mischel1972cognitive}. Non-human primates, including chimpanzees and capuchins, similarly choose larger delayed rewards over immediate ones \cite{rosati2007evolutionary, stevens2005ecology}, suggesting that intertemporal reasoning predates formal economic systems.

Yet real-world investment often depends on others' future actions. Experimental paradigms such as the investment game \cite{berg1995trust}, as well as studies on delayed reciprocity in chimpanzees \cite{melis2006chimpanzees, dufour2007chimpanzee}, demonstrate that both humans and primates can coordinate behavior under deferred, socially contingent outcomes. These findings suggest that the core components of investment cognition—delay tolerance and social projection—are present across species, though rarely stabilized outside human systems.

\section{A Simulateable Architecture for the Emergence of Financial Behavior}

Building on prior work modeling simulateable reciprocity \cite{diau2025cognitivefoundationseconomicexchange}, we propose a lightweight yet extensible architecture for simulating the emergence of financial behavior in multi-agent systems.

While the core behavioral substrate remains grounded in three primitives—partner recognition, reciprocal credence, and cost–return sensitivity—these mechanisms can be implemented in LLM-based agents via structured memory and simple interaction protocols, rather than reinforcement learning. This allows agents to reason about reciprocity and adjust their behavior without engineered rewards or fixed roles.

\begin{itemize}
    \item \textbf{Partner-specific memory:} Each agent maintains a structured record of past interactions indexed by partner identity. These records can include simple counters for cooperative and non-cooperative actions (e.g., "Agent X: 5 helpful, 2 unhelpful").
    
    \item \textbf{Reciprocal evaluation heuristics:} Agents periodically compute a basic reciprocity score for each partner, based on recent cooperative exchanges versus exploitative acts. This score can modulate future willingness to invest in interactions.
    
    \item \textbf{Behavioral updating:} Agents adjust their interaction strategies based on accumulated reciprocity scores, favoring partners with positive balances and withholding cooperation from partners with persistently negative returns.
\end{itemize}

This architecture enables simulation environments to support the spontaneous formation of stable reciprocal networks—without centralized authority, symbolic communication, or predefined roles. It provides a behavioral substrate on which core financial functions can emerge through interaction alone.

\paragraph{Interaction Protocols.}
Agents are not pre-assigned roles or rules. Instead, each agent operates with its own internally defined goals and motivations, making decisions based on prior outcomes and anticipated returns. Interaction choices—whom to engage, when, and why—are driven by internal evaluation of reciprocity history, risk, and projected gain. Cooperative structures emerge not by external design, but from endogenous dynamics shaped by individual strategy updates.

\paragraph{Evaluating Simulation Results via Grounded Behavioral Criteria.}
In multi-agent simulations—especially those involving LLM-based agents—there is often no clear ground truth, and existing evaluations rely on synthetic tasks or arbitrary metrics that lack correspondence to real-world behavior.

We address this by proposing a set of core behavioral criteria—credit, insurance, token exchange, and investment—that reflect well-documented dynamics in actual economic systems. These grounded, interpretable patterns are derived from observable interaction logs and serve as criteria for assessing whether meaningful financial structures have emerged:

\begin{itemize}
    \item \textbf{Credit}: Persistent asymmetric cooperation where one agent contributes without immediate return, followed by delayed reciprocation.
    
    \item \textbf{Insurance}: Clusters of agents that share resources under stochastic harm, exhibiting need-based giving and mutual buffering over time.
    
    \item \textbf{Token-based exchange}: Indirect cooperation mediated through transferable placeholders (e.g., tokens, proxy scores), enabling non-dyadic reciprocity chains.
    
    \item \textbf{Investment}: High-cost actions aimed at uncertain or future-oriented social returns, often involving delayed outcomes contingent on others' behavior.
\end{itemize}

Together, these metrics allow simulations to assess whether decentralized reciprocity stabilizes into distinct, finance-like macrostates.

\section{Implications and Discussion}
\label{others}

\subsection{Limitations of the Framework}

This framework offers a behavioral account of financial structure—grounded in minimal mechanisms of reciprocal interaction rather than institutional design or market optimization. It is not intended as a predictive model, calibrated simulator, or normative design proposal.

We do not model full-scale market dynamics or specify enforcement systems. Rather, we isolate the behavioral preconditions under which finance-like behaviors can emerge spontaneously. Other drivers—such as signaling, imitation, or cultural learning—may also be essential in complex settings. Our aim is not to exclude them, but to show that reciprocity alone suffices to behavioral substrate financial functions. This work provides a foundation upon which more detailed or domain-specific models can build.

\subsection{On the Cognitive Boundary Between Human and Non-Human Reciprocity}
While reciprocity is observed across many species, the extensions required for financial behavior—such as delayed return tracking, risk pooling, symbolic mediation, and long-horizon investment—rarely stabilize in non-human animals. This may reflect more than limited data: it likely signals a cognitive boundary. These behaviors appear to require abstraction, counterfactual reasoning, and sustained partner-specific memory—capacities observed reliably only in humans, and even then emerging gradually during early development. This boundary motivates our approach: to formally reconstruct financial functions from minimal social primitives, scalable beyond the limits of non-human cognition.

\subsection{Future Work}

This framework opens two complementary directions for exploring how financial behavior emerges in the absence of formal institutions.

\paragraph{Agent-based simulation.}  
Multi-agent environments with minimal cognitive assumptions—such as memory, partner recognition, and cost–return sensitivity—can be used to model credit, insurance, trade, and investment. Simulations can help identify the structural conditions under which these behaviors emerge, persist, or fail, and reveal how they interact across different topologies, payoff asymmetries, and risk ecologies.

\paragraph{Behavioral experiments.}
Many financial functions—such as credit and risk pooling—appear even without institutions. Experimental studies in developmental, cross-cultural, and minimal-infrastructure settings can clarify the cognitive substrate of finance by identifying the conditions that support or disrupt these behaviors. Rather than validating models, such work probes what must exist before financial systems can take shape—mapping the true behavioral boundaries of economic cooperation.

\section{Conclusion}
Reciprocity is one of the most ancient and widely observed mechanisms of cooperation—present across primate species and foundational to early human economies. It enables social coordination without formal rules, allowing obligations, support, and exchange to scale through experience and interaction.

This paper shows that reciprocity is not merely a social instinct, but a behavioral substrate from which core financial structures can emerge. Trade—often considered the origin of finance—is reframed as reciprocity in its most symmetric form. Building on this insight, we reconstruct four fundamental financial functions—credit, insurance, token-based exchange, and investment—as structured extensions of the same underlying interaction logic.

By grounding finance in simulateable social behavior, our framework shifts explanatory emphasis away from institutional contracts and enforcement mechanisms, toward the endogenous dynamics of interaction. Financial systems, in this view, do not begin with institutions, but with behavior.

This perspective opens new directions for modeling decentralized cooperation in both human and artificial societies—bridging research in behavioral economics, cognitive science, and multi-agent systems.

\section*{Declaration of LLM Usage}
The authors used OpenAI's ChatGPT to assist in refining phrasing and improving clarity. All theoretical arguments and interpretations are original and authored by the researchers.






\bibliographystyle{abbrvnat}
\bibliography{references}

\end{document}